\documentclass{epl}
\usepackage{graphicx}
\usepackage{bm}
\title{Pairing state in multicomponent superconductors.}
\author{V. Kuznetsova and  V. Barzykin} 
\institute{Department of Physics and Astronomy, University of Tennessee,
Knoxville, TN  37996}


\begin{document}
\maketitle
\begin{abstract}
We use the microscopic weak coupling theory  to predict the pairing state 
in superconductors of cubic, hexagonal, or tetragonal symmetry, where
the order parameter is multicomponent, i.e., transforms
according to either a 2-dimensional or a 3-dimensional
representation of the crystal point group. We show that the 
superconducting phase usually breaks the time-reversal symmetry for singlet multicomponent 
superconductors. The superconducting order parameter for triplet superconductors in most cases 
turns out to be non-magnetic. 
\end{abstract}

 While most superconductors are described by the 
Bardeen-Cooper-Schrieffer (BCS)  weak coupling theory, which 
gives s-wave symmetry of superconducting state,  in many other materials, 
such as $^3$He\cite{osh}, UBe$_{13}$, UPt$_3$\cite{joynt}, the high-Tc cuprates,  Sr$_2$RuO$_4$\cite{RS, Agterberg0, mackenzie},
and PrOs$_4$Sb$_{12}$\cite{Pr}, the pairing state is known to be unconventional. 
(See Refs \cite{gorkov,SU} for a review). If the pairing state in an unconventional superconductor 
or superfluid transforms according to a multi-dimensional representation of the group, the order parameter is also
multicomponent. For example, the $p$-wave order parameter in $^3$He is a $3 \times 3$ complex matrix, i.e.,
has 9 complex components. As a result, many different phases can arise in the 
superfluid state, such as the A-phase\cite{osh}, which is not invariant with respect to rotations 
or time reversal operation, and the B phase, which is both rotationally 
and time-reversal invariant. Unconventional superconductors 
usually have a gapless excitation spectrum, which has important consequences for thermodynamics.  
Instead of describing the fascinating properties of unconventional superconductors, 
we refer the reader to some excellent  books and reviews on the subject\cite{gorkov,SU,joynt,mackenzie,misha,mineev:knizhka}.

  Although the details of the mechanism are essential for strong 
coupling Eliashberg-type calculations, the knowledge 
of the energy spectrum and the coupling constants for quasiparticles 
mediating superconductivity is not required\cite{AGD} in the weak coupling approach, which is adopted below.  
A microscopic description of unconventional 
superconductors can be useful for analysis of experiments in new 
superconducting materials, where unconventional pairing mechanism 
is suspected. The usual suspects are materials with strong Coulomb
correlations, such as the heavy fermions or high-$T_c$ cuprates. In such
materials conventional phonon mechanism gets suppressed\cite{BG}.

The microscopic analysis is somewhat less general then the Ginzburg-Landau (GL) approach of Refs\cite{gorkov,SU} . 
For example, weak coupling 
BCS-type theory always gives the B-phase of superfluid $^3He$\cite{leggett}. 
The non-BCS spin fluctuation feedback effect,  when the interaction 
itself depends upon the superconducting ground state, is required to stabilize the A phase\cite{AB}.
Most, if not all, unconventional superconductors are characterized by the presence of strong Coulomb 
correlations, so the applicability of weak coupling approach for such materials is questionable.
Nevertheless, this approach is often used theoretically as a starting point.

 In what follows we analyze multicomponent superconductors, 
i.e., superconductors belonging to a 2-dimensional or 3-dimensional 
representations of the cubic, hexagonal, and tetragonal crystal point groups,
where the superconducting phase is not uniquely determined by the irreducible representation\cite{gorkov,SU}.
Examples of superconductors where multicomponent state is suspected
are, among others, UPt$_3$\cite{joynt} (hexagonal symmetry),  
Sr$_2$RuO$_4$\cite{RS, Agterberg0}(tetragonal symmetry), and, more recently, PrOs$_4$Sb$_{12}$\cite{Pr}(cubic symmetry)
and Na$_x$CoO$_2 \cdot y$H$_2$O\cite{Baskaran,Maska}.
Like in superfluid $^3He$, many phases can possibly appear in these superconductors, according to 
the GL theory\cite{gorkov,SU}. However, only one of these phases is realized in  the microscopic weak 
coupling theory. The analysis done below is well known\cite{AGD,mineev:knizhka}, and
has been applied to many particular cases, such as the $d$-wave superconductivity in high-T$_c$-s\cite{MBP},
UPt$_3$\cite{haran}, or general multi-band phonon mechanism\cite{ABG}. 

As usual\cite{BG1}, we assume that electrons have a weak short-range attractive
interaction, $U(\bm{r},\bm{r}')$. Since the phonon cutoff frequency is much
less than the Fermi energy, $\omega_D \ll \epsilon_F$, only the quasiparticles
in the vicinity of the Fermi surface participate. Then the interaction 
can be expanded in a complete set of basis functions,
$\chi_{lm}(\bm{p})$, where index $l$ corresponds to different irreducible representations
of the point group, while index $m$ enumerates the basis functions of a given
irreducible representation $l$. This expansion is similar to expansion in 
spherical functions for an isotropic model. The interaction in momentum
space then takes the following form:
\begin{equation}
U(\bm{p},\bm{p}') = \sum_{lm} U_l \chi_{lm}(\bm{p})\chi_{lm}(\bm{p'}),
\label{coupl}
\end{equation}
where $\bm{p}$, $\bm{p'}$ are on the Fermi surface. 
The gap equation for the superconducting order parameter takes
the following form: 
\begin{equation}
\hat{\Delta}_{\alpha \beta, m} (\bm{p}) = |U_l| \chi_{lm}(\bm{p})
\int { d^3 p' \over (2 \pi)^3 } \chi_{lm}(\bm{p'}) \left(T \sum_{\omega_n}
F_{\alpha \beta}(\bm{p'}; i \omega_n) \right),
\label{gap}
\end{equation}
where $F_{\alpha \beta}(\bm{p'}; i \omega_n)$ is the Fourier 
component of anomalous Gor'kov function,
\begin{equation}
\hat{F}_{\alpha \beta}(\bm{r}-\bm{r}', \tau - \tau')
= - \langle T_{\tau} \left(\hat{\Psi}_{\alpha}(\bm{r},\tau)
\hat{\Psi}_{\beta}(\bm{r}',\tau') \right)\rangle,
\label{F}
\end{equation}
and $U_l < 0$ is a constant in Eq.(\ref{coupl}) corresponding to a 
selected pairing channel. 
The two fermion operators inside brackets in Eq.(\ref{F}) anticommute. 
In the presence of the center of inversion, the behavior of
$\chi_l(\bm{p})$ with respect to inversion, $\bm{p} \longrightarrow - \bm{p}$, 
determines the symmetry of $\hat{\Delta}_{\alpha \beta} (\bm{p})$
 - even for singlet, and odd for triplet pairing\cite{gorkov,SU}. The order parameter below $T_c$ has the form:
\begin{eqnarray}
\hat{\Delta}_{\alpha \beta, m} (\bm{p}) & = &
\Delta (T) (i \sigma_y)_{\alpha \beta} \chi_{lm}(\bm{p}), \ \ S = 0, \\
\hat{\Delta}_{\alpha \beta, m} (\bm{p}) & = &
\Delta (T) (\bm{\sigma}_{\alpha \beta} \cdot \bm{d}) \chi_{lm}(\bm{p}), \ \ S = 1.
\end{eqnarray}
Different phases for the multicomponent pairing state were determined from the GL analysis 
in Ref.\cite{gorkov,SU}. The weak coupling approach completely determines the coefficients\cite{AGD} in
the GL functional. The integral gap equation Eq.(\ref{gap}) written in the vicinity of
$T_c$ becomes the GL equation. In what follows we analyze the GL coefficients in degenerate representations of cubic, 
hexagonal, and tetragonal groups to determine possible phases that can arise in the microscopic theory.

\section{The cubic group O$_h$ - two-dimensional representations}

The order parameter for S=0 and S=1 can be written as
\begin{equation}
\hat{\Delta}(\bm{k}) = \Delta (\eta_1 \hat{\varphi}^{(1)}(\bm{k}) + \eta_2 \hat{\varphi}^{(2)}(\bm{k})),
\end{equation} 
where the basis functions are
\begin{eqnarray}
\varphi^{(1)}(\bm{k}) &\propto& k_x^2 + \epsilon k_y^2 + \epsilon^2 k_z^2, \ \ \varphi^{(2)}(\bm{k}) = 
\varphi^{(1)}(\bm{k})^*; \ \ S=0, \\
\vec{\varphi}^{(1)}(\bm{k}) &\propto& \hat{x} k_x + \epsilon \hat{y} k_y + \epsilon^2 \hat{z} k_z, \ 
\ \vec{\varphi}^{(2)}(\bm{k}) = \vec{\varphi}^{(1)}(\bm{k})^*; \ \ S=1.
\end{eqnarray}
Here $\epsilon = \exp{(2 \pi i/3)}$, and the wave functions are normalized to 1:
\begin{equation}
\langle | \hat{\varphi}^{(i)}(\bm{k}) |^2 \rangle_{F.S.} = 1, \ \ |\eta_1|^2 + |\eta_2|^2 = 1.
\end{equation}
For singlet order parameter, we find:
\begin{equation}
F = N_0 \Delta^2 \ln{\left[\frac{T}{T_c}\,\right]} (|\eta_1|^2 + |\eta_2|^2) + \alpha N_0 \Delta^4 \frac{5 \zeta(3)}{16 \pi^2 T_c^2}\, 
(|\eta_1|^4 + |\eta_2|^4 + 4 |\eta_1|^2 |\eta_2|^2).
\label{cub2D}
\end{equation}
Here $N_0$ is the density of states for one spin direction, $\alpha$ is a coefficient determined from the 
Fermi surface averages:
\begin{equation}
\alpha = \frac{\langle | \hat{\varphi}^{(1)}(\bm{k}) |^4 \rangle_{F.S.}}{\langle | \hat{\varphi}^{(1)}(\bm{k}) |^2 \rangle_{F.S.}^2}\,.
\end{equation}
For the isotropic spherical shape of the Fermi surface we find that $\alpha = 10/7$. While $\alpha$ will be different
for an arbitrary Fermi surface, \textit{the form} of the functional Eq.(\ref{cub2D}) 
is completely determined by  the cubic symmetry, since certain averages, such as $\langle \hat{\varphi}^{(1)}(\bm{k})^4 \rangle_{FS}$, 
must vanish by symmetry (in this case, rotation around the 3-fold axis). 
This functional leads to the phase $(1, 0)$ with a magnetic d-wave order parameter, 
\begin{equation}
\hat{\Delta}(\bm{k}) \propto k_x^2 + \epsilon k_y^2 + \epsilon^2 k_z^2,
\end{equation}
which breaks $T \rightarrow -T$ symmetry,
and has point nodes along the diagonals of the cube. Note that in this case higher order terms are not necessary to determine the phase.
In the triplet case, we obtain, for the spherical Fermi surface:
\begin{equation}
F^{(4)-(6)} =  \frac{7 \zeta(3) N_0 \Delta^4}{10 \pi^2 T_c^2}\, 
(|\eta_1|^4 + |\eta_2|^4 + |\eta_1|^2 |\eta_2|^2) - 
\frac{93 \zeta(5) N_0 \Delta^6}{2240 (\pi T_c)^4} (\eta_1^3 \eta_2^{*3} + \eta_1^{*3} \eta_2^3)
\end{equation}
The 6-order term is necessary in this case to distinguish between the two non-magnetic superconducting phases. The magnetic class
turns out to be the trivial $D_4 \otimes R$, with the order parameter
\begin{equation}
\bm{d}(\bm{k}) \propto 2 \hat{x} k_x - \hat{y} k_y - \hat{z} k_z.
\end{equation}
This order parameter is nodeless, i.e., thermodynamic properties of this state are exponential at low temperature, similar
to BCS s-wave. For a general Fermi surface we found that the sign of the 6-th order term is always negative. However, the coefficient
$\kappa$ in the 4-th order term, 
\begin{equation}
F^{(4)} \propto |\eta_1|^4 + |\eta_2|^4 + \kappa  |\eta_1|^2 |\eta_2|^2
\end{equation}
depends on the shape of the Fermi surface. We find that
$- 0.8 \le \kappa \le 4$, which makes magnetic class $O(D_2)$ ($\kappa > 2$) also a possibility.

\section{ Three-dimensional representations}

For the four 3D vector representations, the basis functions can be chosen in the following form:
\begin{eqnarray}
F_{1g}(S=0): & & k_y k_z (k_y^2 - k_z^2), \ \ k_z k_x (k_z^2 - k_x^2), \ \ k_y k_x (k_x^2 - k_y^2);  \\
F_{1u}(S=1): & & \hat{y}k_z - \hat{z}k_y, \ \ \hat{z} k_x - \hat{x} k_z, \ \ \hat{x} k_y - \hat{y} k_x; \\
F_{2g}(S=0): & & k_y k_z, \ \ k_x k_z, \ \ k_x k_y; \\
F_{2g}(S=1): & & \hat{y} k_z + \hat{z} k_y, \ \ \hat{z} k_x + \hat{x} k_z, \ \ \hat{x} k_y + \hat{y} k_x. 
\end{eqnarray}
The Free energy can be written in the universal form:
\begin{equation}
F = N_0 \Delta^2 \ln{\left[\frac{T}{T_c}\,\right]} (\vec{\eta} \cdot \vec{\eta*}) + 
\beta_1 N_0 \Delta^4 (\vec{\eta} \cdot \vec{\eta*})^2 +
\beta_2 N_0 \Delta^4 |\vec{\eta} \cdot \vec{\eta}|^2 + \beta_3 N_0 \Delta^4 \sum_i |\eta_i|^4 
\label{GLE}
\end{equation}

\noindent
a) \underline{$F_{1g}$.}  We find the following GL coefficients in Eq.(\ref{GLE}) for a spherical Fermi surface:
\begin{equation}
\beta_1 = \frac{2205 \zeta(3)}{3536 \pi^2 T_c^2}\,, \ \ \beta_2 = \frac{\beta_1}{2}\,, \ \ \beta_3 = \frac{9}{22}\, \beta_1,
\end{equation}
which gives magnetic superconducting class $D_3(E)$, with a phase $(1, \epsilon, \epsilon^2)$, 
given in terms of the above basis functions. This phase is ferromagnetic, with orbital magnetic moment pointing along the 3-fold
axis of the cube. The order parameter has 2 point nodes, at the points of intersection of the spontaneous 3-fold anisotropy
axis with the Fermi surface. For a general Fermi surface, the phase is not completely determined. We found that
\begin{equation}
\beta_1 > 0, \ \ \frac{\beta_2}{\beta_1}\, = \frac{1}{2}\,, \ \ \frac{\beta_3}{\beta_1}\, \ge - \frac{3}{2}\,.
\end{equation}
This gives a line on more general phase space for the GL theory. Thus, magnetic phases $D_3(E)$, $D_4(E)$, and
a non-magnetic phase $D_4(C_4) \otimes R$ are possible depending on the shape of the Fermi surface.

\noindent
b) \underline{$F_{2g}$.}
For a spherical Fermi surface, we find:
\begin{equation}
\beta_1 = \frac{5 \zeta(3)}{8 \pi^2 T_c^2}\,, \ \ \beta_2 = \frac{\beta_1}{2}\,, \ \ \beta_3 = 0.
\label{cubeR}
\end{equation}
This case falls on the boundary between phases $(1, \epsilon, \epsilon^2)$ and $(1, i, 0)$, with
accidental degeneracy between $D_4(E)$ and $D_3(E)$.  For an arbitrary Fermi surface, we find:
\begin{equation}
\frac{\beta_2}{\beta_1}\, = \frac{1}{4}\,, \ \ \beta_2 > 0, \ \ \frac{\beta_3}{\beta_1}\, \ge - \frac{5}{4}\,,
\end{equation}
and the magnetic phases  $D_3(E)$, $D_4(E)$, and non-magnetic $D_4^{(2)}(D_2) \otimes R$ are possible.

\noindent
c) \underline{$F_{1u}$ and $F_{2u}$.}  For both triplet representations we find, in case of spherical Fermi surface:
\begin{equation}
\beta_1 = \frac{63 \zeta(3)}{80 \pi^2 T_c^2}\,, \ \ \beta_2 = - \frac{\beta_1}{3}\,, \ \ \beta_3 = 0,
\end{equation}
which falls on the boundary between 2 non-magnetic phases: $(1,1,1)$ 
(class $D_3(C_3) \otimes R$ for $F_{1u}$, $D_3 \otimes R$ for $F_{2u}$)
and $(1, 0, 0)$ (class $D_4(C_4) \otimes R$ for $F_{1u}$, class $D_4^{(2)}(D_2) \otimes R$ for $F_{2u}$).
For the general Fermi surface one of these non-magnetic phases will be realized. We find that
\begin{equation}
\frac{\beta_3}{\beta_1}\, + 3 \frac{\beta_2}{\beta_1}\, + 1 = 0, \ \ \beta_1 \ge 0, \ \ \frac{\beta_2}{\beta_1}\, < 0.
\end{equation}

\section{The hexagonal group D$_{6h}$}

Multicomponent order is possible in hexagonal group, if the pairing wave function transforms according to one
of the 2D irreducible representations.  For $E_1$, the GL functional is given by Eq.(\ref{GLE}), with the 
order parameter now a 2D vector:
\begin{eqnarray}
E_{1g} (S=0): & & \psi(\bm{k}) \propto \eta_x k_z k_x + \eta_y k_z k_y  \\
E_{1u} (S=1): & & \bm{d}(\bm{k}) \propto \eta_x \hat{z} k_x + \eta_y \hat{z} k_y
\end{eqnarray} 
For an arbitrary Fermi surface we find:
\begin{equation}
\frac{\beta_2}{\beta_1}\, = \frac{1}{2}\,, \ \ \beta_1 > 0, \ \ \beta_3 = 0
\end{equation}
This gives rise to the phase $(1, i)$, in both
singlet and triplet cases, or the symmetry class $D_6(E)$, which breaks the time-reversal symmetry, and possesses ferromagnetism.
For the particular case of a spherical or cylindrical Fermi surface we find:
\begin{equation}
\beta_1 (sph.) = \frac{5 \zeta(3)}{8 \pi^2 T_c^2}\,, \ \ \beta_1 (cyl.) = \frac{7 \zeta(3)}{16 \pi^2 T_c^2}\,.
\end{equation}

For $E_{2g}$ the GL functional is the same as for the 2D representation $E_g$ in the cubic lattice. The basis
functions are, of course, not the same:
\begin{equation}
E_{2g} (S=0): \ \ \varphi^{(1)}(\bm{k}) \propto (k_x + i k_y)^2, \ \ \varphi^{(2)}(\bm{k}) =  \varphi^{(1)}(\bm{k})^*;
\end{equation}
For an arbitrary Fermi surface the form of the functional, Eq.(\ref{cub2D}), stays the same, while the coefficient $\alpha$ 
depends on Fermi surface averages of the basis function. For example, $\alpha$ for the particular cases
of spherical or cylindrical Fermi surface given by:
\begin{equation}
\alpha(sph) = \frac{10}{7}\,, \ \ \alpha(cyl) = 1.
\end{equation}
The resulting phase $\psi(\bm{k}) \propto  (k_x + i k_y)^2$   belongs to the symmetry class $D_6(C_2)$, and
is an orbital ferromagnet.
Finally, for $E_{2u}$ triplet representation, with basis wave functions
\begin{equation}
E_{2u} (S=1): \ \  \varphi^{(1)}(\bm{k}) \propto (\hat{x} + i \hat{y})(k_x + i k_y), \ \ \varphi^{(2)}(\bm{k}) =  \varphi^{(1)}(\bm{k})^*
\end{equation}
the free energy functional is given by:
\begin{equation}
F = \ln{\left[\frac{T}{T_c}\,\right]} N_0 \Delta^2 (|\eta_1|^2 + |\eta_2|^2) + \alpha N_0 \Delta^4 \frac{7 \zeta(3)}{16 \pi^2 T_c^2}\, (|\eta_1|^4 + |\eta_2|^4),
\end{equation}
which gives either the p-wave phase $ \hat{x} k_x - \hat{y} k_y$ (symmetry class  $D_2 \otimes R$), or 
$\hat{x} k_y + \hat{y} k_x$ (symmetry class $D_2(C_2) \otimes R$). Unfortunately, BCS does not distinguish between the two
phases in the 6-th order. Here $\alpha > 0$ is a coefficient, which depends on the shape of the Fermi surface. For a cylindrical
and spherical Fermi surface,
\begin{equation}
\alpha(sph) = \frac{6}{5}\,, \ \ \alpha(cyl) = 1.
\end{equation}

\section{The tetragonal group D$_{4h}$.}

Since the basis functions for the $2D$ representations $E_u$ and $E_g$ of the tetragonal group are exactly the same as for 
hexagonal $E_{1u}$ and $E_{1g}$ representations, not surprisingly, the GL functionals and the resulting phases 
also turn out to be identical. The class in tetragonal group is $D_4(E)$, which allows ferromagnetism along $z$-axis.
There are 2 point nodes for the singlet, and a line of nodes in the triplet case. 

To summarize, we have considered phases realized in microscopic weak coupling theory in 
multicomponent superconductors. We found that, for the most part, weak coupling mechanism in superconductors
selects the most interesting \textbf{magnetic phase}, which is similar to $A$-phase in superfluid $^3He$.
In this pairing state the time-reversal symmetry is broken. The properties of such superconductors 
are well known\cite{gorkov,SU,misha,mineev:knizhka}. Similar to bulk 
ferromagnets, a pairing state which breaks time-reversal symmetry tends to form domain 
structure, with magnetic moment arising from circulating electric current 
in the domain walls. Thus, there is no net magnetic moment in the bulk. 
Perhaps, the most interesting phenomena for multicomponent superconductors arise in
magnetic field. An obvious consequence of magnetic field is
the change of phase for a certain direction, which has to do with the symmetry\cite{misha,mineev:knizhka}.
This gives rise to very unusual topological excitations and vortex lattices\cite{misha,Agterberg,Pasha},
and an $H-T$ phase diagram, which reflects the possibility of many different phases\cite{Pr,adenwalla}
or an abrupt change of slope for $H_{c2}(T)$\cite{Maska}. Another detectable property is the splitting 
of $T_c$ under uniaxial stress\cite{UPT3}.

Our results are summarized in table I, where we also list specific heat jump at 
$T_c$ for simple Fermi surfaces, such as cylindrical or spherical, written as a ratio to s-wave BCS theory specific heat jump,
$(\Delta C/C) = 12/7 \zeta(3) \simeq 1.42$. Our results for singlet superconductors
are consistent with Ref. \cite{ABG}, where the BCS mechanism for multi-pocket Fermi surfaces was considered.  
The main conclusion of Ref. \cite{ABG} was that it is the peculiar shape of the Fermi surface, such as its
multi-pocketed structure,  which determines the phase realized in BCS approach for a multicomponent order parameter
and leads to a magnetic phase. We found that this statement is only partially correct. For all degenerate 
representations of the cubic group, except for the 2D representation $E_g$, the superconducting phase 
depends on the shape of the Fermi surface. The weak coupling approach, however, introduces certain constraints on 
the GL coefficients and possible phases. For the $2D$ representations of the hexagonal and tetragonal groups 
and the representation $E_g$ of the cubic group, the magnetic superconducting phase in the weak coupling approach is 
determined by \textbf{symmetry alone}.  
The shape of the Fermi surface determines
the specific heat jump at $T_c$ and even the form of the gap function, since the basis functions
for irreducible representations of point groups are not unique, but not the superconducting phase. 
The only requirement for magnetic phase to appear is
that the superconducting order parameter transforms according to one of these irreducible representations. 

This work was supported by TAML at the University of Tennessee.

\begin{table}
\caption{Magnetic classes obtained in weak coupling approach for degenerate representations of
cubic, tetragonal, and hexagonal point groups. The basis functions 
are given in text. $\epsilon \equiv \exp{(2 \pi i/3)}$}
\begin{center}
\begin{tabular}{|c|c|c|c|c|c|c|}
\hline \hline
Symmetry & irr. & Symm. &  $(\eta_1, \eta_2)$ or & Magnetic & \multicolumn{2}{|c|}{ $\Delta C/C$, simple FS }\\
\cline{6-7}
group & repr. &  class & $(\eta_1, \eta_2, \eta_3)$ & properties &  spher. & cyl. \\
\hline
      & $E_g$     & $O(D_2)$             & $(1,0)$                     & AFM       & $0.7$ BCS        & N/A \\
      & $E_u$     & $D_4 \otimes R$      & $(1,1)$                     & non-magn. & $5/6$ BCS        & N/A \\
      &           & $O(D_2)$             & $(1,0)$                     & AFM       & N/A              & N/A \\
      & $F_{1g} $ & $D_3(E)$             & $(1, \epsilon, \epsilon^2)$ & FM        & $\simeq 0.5$ BCS & N/A \\
$O_h$ &           & $D_4(E)$             & $(1, i, 0)$                 & FM        &  N/A             & N/A \\
      &           & $D_4(C_4) \otimes R$ & $(1,0,0)$                   & non-magn. &  N/A             & N/A \\
      & $F_{1u}$  & $D_3(C_3) \otimes R$ & $(1,1,1)$                   & non-magn. & $5/6$ BCS        & N/A \\
      &           & $D_4(C_4) \otimes R$ & $(1,0,0)$                   & non-magn. & $5/6$ BCS        & N/A \\
Cubic & $F_{2g} $ & $D_3(E)$             & $(1, \epsilon, \epsilon^2)$ & FM        & $0.7$ BCS        & N/A \\
      &           & $D_4(E)$             & $(1, i, 0)$                 & FM        & $0.7$ BCS        & N/A \\
      &     & $D_4^{(2)}(D_2) \otimes R$ & $(1,0,0)$                   & non-magn. & N/A              &  N/A \\
      & $F_{2u}$  & $D_3 \otimes R$      & $(1,1,1)$                   & non-magn. & $5/6$ BCS        & N/A \\
      &           & $D_4(D_2) \otimes R$ & $(1,0,0)$                   & non-magn. & $5/6$ BCS        & N/A \\
\hline
          & $E_{1g} $ & $D_6(E)$             & $(1,i)$  & FM        & $0.7$ BCS & BCS \\ 
$D_{6h}$  & $E_{1u}$  & $D_6(E)$             & $(1,i)$  & FM        & $5/6$ BCS & BCS\\
          & $E_{2g} $ & $D_6(C_2)$           & $(1,0)$  & FM        & $0.7$ BCS & BCS\\
Hexagonal & $E_{2u}$  &  $D_2 \otimes R$     & $(1,1)$  & non-magn. & $5/6$ BCS & BCS\\
          &           &  $D_2(C_2) \otimes R$& $(1,-1)$ & non-magn. & $5/6$ BCS & BCS\\
\hline
$D_{4h}$   & $E_g $ & $D_4(E)$ & $(1,i)$ & FM & $0.7$ BCS & BCS\\
Tetragonal & $E_u$  & $D_4(E)$ & $(1,i)$ & FM & $5/6$ BCS & BCS \\
\hline \hline
\end{tabular}
\end{center}
\end{table}

\end{document}